\documentclass[runningheads]{llncs}
\usepackage{graphicx}
\usepackage{float}
\usepackage{lscape}
\usepackage {xcolor}
\usepackage[colorlinks=true, allcolors=blue]{hyperref}

\usepackage{comment}
\usepackage[normalem]{ulem}
\usepackage{tcolorbox}
\usepackage{booktabs}

\begin{document}
\title{LLM-based agents for automating the enhancement of user story quality: An early report}
\author{Zheying Zhang\inst{1}\orcidID{0000-0002-6205-4210} \and
Maruf Rayhan\inst{1} \and
Tomas Herda\inst{2}\orcidID{0009-0005-2912-380X} \and Manuel Goisauf\inst{2} \and Pekka Abrahamsson\inst{1}\orcidID{0000--0002-4360-2226}}

\authorrunning{Z. Zhang et al.}
\institute{Tampere University, Finland\\
\email{\{zheying.zhang, maruf.rayhan, pekka.abrahamsson\}@tuni.fi}\and
Austrian Post, Rochusplatz 1, 1030 Vienna, Austria
\email{\{tomas.herda,manuel.goisauf\}@post.at}}

\maketitle              % typeset the header of the contribution
\begin{abstract}
In agile software development, maintaining high-quality user stories is crucial, but also challenging. This study explores the use of large language models (LLMs) to automatically improve the user story quality in Austrian Post Group IT agile teams. We developed a reference model for an Autonomous LLM-based Agent System (ALAS), and implemented it at Austrian Post Group IT. The quality of use stories in the study and the effectiveness of these agents were assessed by 11 participants across six agile teams. Our findings demonstrate the potential of LLMs in improving user story quality, contributing to the research on AI’s role in Agile
development, and providing a practical example of the transformative impact of AI in an industry setting.

\end{abstract}
%
%
%
%\section{First Section}
\section{Introduction}

Effective requirements management is critical in software projects, ensuring that the final product meets customer needs and business goals to deliver value \cite{wiegers2013software}\cite{6146379}. In agile software projects, requirements are iteratively specified and prioritized, typically as user stories, allowing for responsiveness to evolving user needs and ensuring value delivery in iterative and incremental cycles. The quality of user stories \cite{6146379}\cite{glinz2020handbook}\cite{ferreira2022towards}\cite{dalpiaz2019detecting}\cite{lucassen2016improving} directly influences the development cycle's velocity and the fulfillment of customer expectations. However, ensuring the completeness, consistency, unambiguity, testability, etc. of user stories, i.e. good user stories, presents challenges. 

As agile methodologies emphasize rapid iteration and adaptability, the potential of large language models (LLMs) to assist in user story analysis is becoming increasingly significant. The advanced natural language processing capabilities of LLMs present a promising solution for automating and enhancing user story quality. By analyzing, refining, and even generating user stories, LLMs can provide substantive assistance to product owners, developers, test engineers, etc. in requirements management for software development.

This research aims to explore the potential of automating user story quality enhancement by integrating LLM agents into real-world agile software development environments. To achieve this goal, we propose a reference model for LLM agent systems, based on which we implement and evaluate the role of agents in improving the quality of user stories. This paper presents the preliminary results of deploying the system at Austrian Post Group IT, with a particular focus on the quality improvement of user stories in the company’s mobile delivery project. We evaluate the agents' effectiveness in six agile teams across the company. The results contribute to the emerging discussion around AI’s role in agile software development, demonstrating a proof of concept of the transformative impact of the LLM in assisting with industry-demanding tasks.

\section{User Story Quality}
%\subsection{User story quality}
In agile software projects, requirements are often expressed as user stories \cite{cohn2004user}, which are brief descriptions of functionalities or features from the user's perspective, emphasizing their needs and the value the feature brings. A widely accepted template for user stories is: "As a [role], I want [requirement] so that [benefit]." This effectively includes the core elements such as the intended user (role), the desired system functionality (requirement), and, optionally, the underlying rationale (benefits). Additionally, every user story should be accompanied by a set of acceptance criteria (AC) that outline detailed conditions a user story must meet to be considered complete and acceptable, including functional behavior, business rules, and quality aspects to be tested. The AC makes a user story more concrete and less ambiguous \cite{cohn2004user}.

Writing good user stories is essential in software projects, as they convey the needs and perspectives of users and guide the development team in implementing the expected functionalities. Beyond general guidelines for quality in requirements engineering, such as ISO/IEC/IEEE 29148-2011 \cite{6146379} and IREB guidelines \cite{glinz2020handbook}, various frameworks include a set of criteria for assessing the quality of user stories. For example, the INVEST framework \cite{INVEST} includes attributes such as independence, negotiability, value, estimability, small, and testability, thereby promoting practical and well-defined requirements. The Quality User Story (QUS) framework \cite{lucassen2015forging} evaluates user stories based on their syntactic, semantic, and pragmatic qualities, including criteria such as well-formedness, atomicity, minimalism, conceptual soundness, problem-orientation, unambiguity, completeness, and uniqueness. These frameworks include a variety of criteria for high-quality user stories. Regardless of their diversity, they adhere to industry standards \cite{6146379}\cite{glinz2020handbook} that ensure user stories are concise, clear, and achievable, and contribute to the success of software development projects and positive user experiences.

Despite the widespread adoption of user stories and available criteria for good user stories, the methods for assessing and enhancing their quality are still relatively limited. Berry et al.\cite{berry2006new} introduced a quality model and a prototype tool named QuARS \cite{fabbrini2001linguistic} for automatic linguistic analysis of natural language requirements. Lucassen et al.\cite{lucassen2015forging}\cite{lucassen2016improving} proposed the QUS framework and the AQUSA software tool, which employs natural language processing (NLP) techniques to detect quality defects in user stories, focusing on syntax and pragmatics.

Recent research has increasingly focused on leveraging LLMs to assist in software engineering tasks \cite{nguyen2023generative}. In requirements engineering, research has focused on utilizing LLMs for requirements elicitation, analysis, and classification, and provided preliminary empirical evidence of their significant impact on requirements engineering tasks. For example, White et al. \cite{white2023chatgpt} introduced a catalog of prompt patterns for stakeholders to interactively evaluate the completeness and accuracy of software requirements. Ronanki et al. \cite{ronanki2023investigating} conducted a comparative analysis between ChatGPT-generated requirements and those specified by requirements experts from both academia and industry. The results revealed that LLM-generated requirements, while abstract, are consistently understandable. This indicates the potential of LLMs, like ChatGPT, in automating various tasks through its NLP capabilities. Moreover, the advent of LLMs spurred the advancement of prompt techniques. A prompt is an input given to a language model to guide its response generation. It is a basic tool for optimizing model performance for specific tasks by structuring interactions to produce desired outcomes. Researchers have explored various prompt patterns to maximize these benefits, including the commonly used direct questions and instructional prompts. Specifically, the chain-of-thought \cite{wei2022chain} encourages models to articulate intermediate steps to enhance their performance on complex reasoning tasks. The k-shot prompting \cite{brown2020language} incorporates illustrative examples to enhance the model's task understanding and expected behavior. The fact checklist \cite{white2023chatgpt} pattern ensures the model addresses or verifies specific criteria in its responses. 

While interest in applying LLMs to engineering tasks is growing, research on their industrial implementation and performance evaluation remains limited. This gap forms the goal of our study, which aims to connect the theoretical potential of LLMs with practical application to gather feedback from industrial professionals. 

\section{Methodology}
To effectively apply LLMs to engineering tasks, we propose a reference model of LLM-based agents. This model forms a framework for designing an Autonomous LLM-based Agent System (ALAS), which we elaborate on in this section.

\subsection{A Reference Model of LLM-based Agents}

\begin{figure}[ht]
\centering
\includegraphics [scale=0.45]{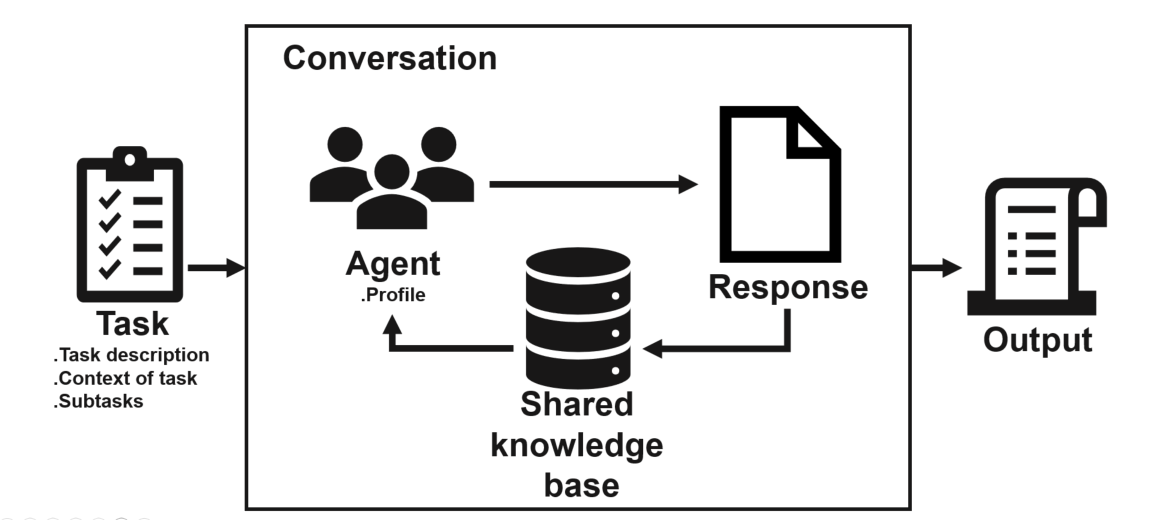}
\caption{A reference model of an agent system}
\label{fig:referencemodel}
\end{figure}

The model depicted in Fig. \ref{fig:referencemodel} conceptualizes the interaction among LLM agents in completing a task. The model is composed of basic constructs: task, agent, shared knowledge base, and response. Together, they form a framework that facilitates the generation of the desired output to fulfill the task’s objectives.

The \textit{task} initiates the interaction and contains inputs that define the work scope and objectives. It can vary in nature, such as a coding task or a requirements review task in software engineering. The task articulates a comprehensive description, contextual information for task understanding and execution, and the expected outcome. It may also prescribe procedural steps for agents to follow in producing the desired output.  
\vspace{-15pt} 
\vspace{-15pt}

Each \textit{agent}, i.e. Agent i, represents an instance of an LLM model, possessing a profile of a unique role in performing the task. For example, Agent PO (Product Owner) represents a product owner role in an agile project, with a focus on the successful delivery of a product that meets the users' needs; while Agent RE (Requirements Engineering) represents a requirements engineer role, and specializes in analyzing and improving the requirements quality, ensuring requirements are compliance with standards like ISO/IEC/IEEE 29148-2011 \cite{6146379}.  

An agent generates an output, called a \textit{response}, based on the task description. The response is then added to the \textit{shared knowledge base},  a repository that contains information on the task and the conversation history. It is initiated with the task description and the first step in the conversation and continues to expand with the responses generated by agents. It provides a dynamic resource that informs subsequent agents of the context of the conversation and the next step towards completing the task, enabling informed transitions from one agent to the next and maintaining a coherent dialogue.

A \textit{conversation} takes place among participating agents. It begins with establishing the context and requirements through a task description. Agents, using the shared knowledge base, engage in a collaborative conversation, each contributing their expertise towards the task completion. The final output is an accumulation of these collaborative efforts, embodying the collective intelligence of the participating agents.

The model provides a structured approach to integrating LLMs in the requirements engineering process, thereby assisting agile teams in developing high-quality software products.

\subsection{Implementing an Agent Environment}

An agent-based system’s strength lies in the AI agents’ ability to communicate and execute tasks, thereby facilitating the automation of software development tasks. The implementation of such a system, as described in the reference model, is a pivotal step in harnessing LLMs to assist software development practices. Our Autonomous LLM-based Agent System (ALAS)  was designed to automate AI agents’ collaboration across various software development scenarios. 

In ALAS, agents are powered by LLMs, and their collaboration is orchestrated through prompts. These prompts define the actions every agent is expected to perform at each step. There are two categories of prompts in ALAS: initial prompts (\textit{Prompt\textsubscript{i}, 1$\le$i$\le$k}), which prepare \textit{k} participating agents for their task responsibilities,  and follow-up prompts (\textit{Prompt\textsubscript{i}, i$>$k}), which are dynamically constructed to guide agents through the necessary steps for interaction and successful task completion.

\begin{center}    
\textbf{\textit{Prompt\textsubscript{i} = Profile\textsubscript{i} + Task + Context of task + Subtask\textsubscript{i}}}, \textit{(1$\le$i$\le$k)}\\
\textbf{\textit{Prompt\textsubscript{i} = Subtask\textsubscript{i} + Response\textsubscript{i-1}}}, \textit{(i$>$k)}\\
\end{center}

{\footnotesize\noindent where
 \textit{Prompt\textsubscript{i}}: 1$\le$i$\le$k: first prompt to \textit{Agent\textsubscript{i}}, with \textit{k} agents engaged in completing the Task; i$>$k: Prompt for Subtask i; \textit{Profile\textsubscript{i}}: \textit{Agent\textsubscript{i}'s profile}; \textit{Task}: Task to complete;
  \textit{Context of Task}: Background information where the task is situated; 
\textit{Subtask\textsubscript{i}}: Subtask i;
\textit{Response\textsubscript{i-1}}: Response produced after completing Subtask\textsubscript{i-1}.}\\

The prompts ensure that each agent has a clear understanding of their role and the steps they need to take to contribute to the completion of the overall task. After the initial "icebreaking" phase, where agents get acquainted with the task and their roles, subsequent prompts are tailored based on the responses to the prior subtasks to maintain a coherent dialogue flow. Consequently, the implementation of our system includes two phases: task preparation and task conduction. An example of the two phases is illustrated in Fig. \ref{fig:AI plan}.

\subsubsection{Task Preparation Phase}
The task preparation phase aims at formulating initial prompts, enabling agents to gain a comprehensive understanding of their roles and expected contributions to the task. This phase establishes the groundwork for agent interaction and task execution. It involves defining the task, articulating its context, specifying agents’ roles and responsibilities, and planning their interaction sequence which simulates the communication dynamics among agile team members to complete a task in real-life projects. This is an iterative process to formulate and optimize prompts to ensure agents communicate effectively to produce the desired output. Various prompt patterns and techniques can be employed for this purpose.

\textit{The persona pattern}\cite{white2023chatgpt} involves creating a \textit{Profile\textsubscript{i}} for each agent, guiding it to adopt a specific character or role. This pattern shapes the agent’s responses to reflect the required expertise and perspective.

\textit{The k-shot prompt}\cite{brown2020language} is a technique for in-context learning. It provides explicit instructions or examples of the desired output. and is particularly useful in formulating the \textit{Task} description and the \textit{context of task}. For example, to generate a product vision statement, the prompt may include a description of elements of a vision statement or an example from another product, serving as a one-shot prompt for creating a vision statement for our product.

\textit{The AI planning}\cite{silver2022pddl} helps to generate plans for completing a task by breaking the task into smaller and manageable \textit{Subtask\textsubscript{i}}, and assigning responsible agents for each subtask. 

\textit{The fact check list pattern}\cite{white2023chatgpt} can be applied for verifying and validating LLM’s outputs. This pattern instructs a model to create key facts in its output. This checklist can help identify any potential errors or inconsistencies in the output. Typically, it is used in conjunction with other prompt patterns, such as the AI plan pattern, to ensure that the model generates a relevant plan, complying with the task description. It also helps to assess the output's accuracy and the supporting evidence.

\textit{The Chain of Thought (CoT)}\cite{wei2022chain}\cite{zhang2022automatic} method guides LLMs through a step-by-step reasoning process to answer questions. It helps clarify the model's thought process and the rationale behind its conclusions.

It is worth noting that the patterns and techniques presented here are not exhaustive, but rather a selection of those commonly used for task preparation. The appropriate choice of patterns and techniques is crucial to effectively support the task preparation phase, guaranteeing that prompts are effectively formulated and optimized to facilitate successful task completion.

\subsubsection{Task Conduction Phase}

In the task conduction phase, agents dynamically interact to execute subtasks, using prompts to guide their activities. In practice, this is an iterative and incremental process, like what an agile team performs in a software development project. Each agent sequentially engages with the subtasks to execute their responsibility by following the structured prompt. The use of the previous response in the current prompt ensures that each agent's response is relevant and builds upon the previous work. This iterative collaboration is like the daily stand-ups and sprint reviews in Scrum, where each team member's work is informed by the overall sprint progress. At the same time, the prompt structure ensures that the task evolves dynamically with each agent's previous response, reflecting the adaptive and responsive nature of an agile project where plans and tasks are continuously refined based on ongoing feedback and developments. The final output is incrementally generated based on the agents' responses.

\section{Experiments}

Following the implementation of ALAS, we evaluated its effectiveness in improving user story quality within agile teams at Austrian Post Group IT with a robust agile framework. The company has multiple teams, working synchronously across numerous systems and applications orchestrated within Agile Release Trains \cite{austriaPost}. User stories play an important role in planning and prioritizing the implementation of these systems, facilitating communication and collaboration across diverse teams. High-quality user stories are essential for successful development projects. Recognizing this criticality, we explored the potential of ALAS to enhance user story quality. 

\begin{figure}[!ht]
\centering
\includegraphics[scale=0.45]{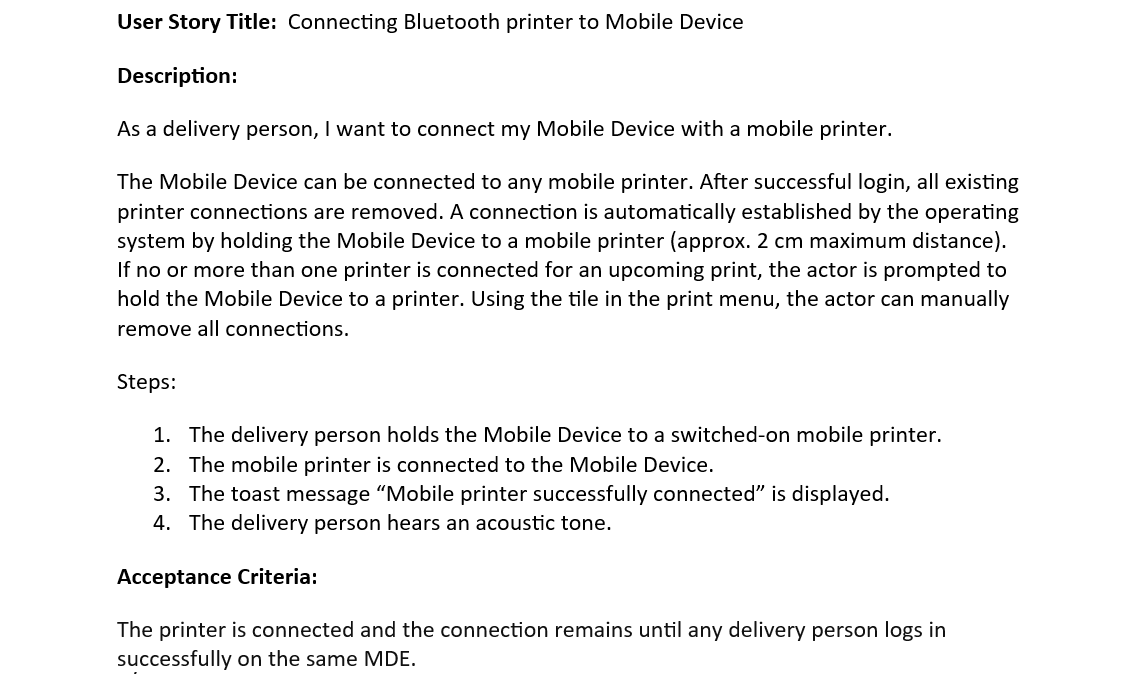}
\caption{User story 1 (US1) - a user story example in the Mobile Delivery project}
\label{fig:userstoryexample}
\end{figure}

Our experiments evaluate the effectiveness, benefits, concerns, and overall satisfaction with the quality of user stories improved by the ALAS. %our LLM-Agent application. 
We evaluate our implementation using 25 synthetic user stories for a mobile delivery application - a tool that helps postal workers prepare and deliver %, and account for 
their parcels%deliveries
. An example user story is shown in Fig. \ref{fig:userstoryexample}, which describes a delivery person's task of synchronizing a mobile device with a mobile printer.

\vspace{-15pt}

\subsection{Setting up Experiments}

The experimental setup, i.e. task preparation phase, involves articulating a task alongside its context, defining the agents' profiles, and planning subtasks. This setup is an iterative process of creating and refining prompts. We applied different prompting techniques and patterns to maximize the capabilities of the agents. 

\subsubsection{Task and Context of Task}
The task %in the experiments 
was to improve the quality of user stories and ensure alignment with the organizational standards for requirements engineering. These user stories, originally from the Mobile Delivery project, require enhancement not only in clarity, completeness, correctness, consistency, etc. but also in their relevance to the overall functionalities of the application, aligning with the business objectives. To facilitate this, we added two documents when describing the task. One is a minimum viable product (MVP) document that details the basic features of the mobile delivery application. It serves as a blueprint to guide agents in refining user stories in a way that resonates with core product features. Another is a product vision statement, structured using the NABC (Needs, Approach, Benefit, and Competition) value proposition template \cite{NABC}. This document provides a strategic overview of the application, covering aspects such as the client’s needs, the solution, client benefits, and unique value propositions. The combination of these two documents provides agents with a comprehensive background so that they can execute their tasks with a sufficient understanding of both the technical and strategic objectives of the project. 

\subsubsection{Agent profiles}

In the process of user story quality analysis and improvement within an agile framework, key roles such as product owner, requirements engineer, quality assurance specialists, and agile team members are often involved. Each role contributes unique skills to ensure that user stories are not only technically clear but also align with the project's goals and users' needs. To simplify our experimental setup, Austrian Post Group IT identified two main focus roles: product owner and requirements engineer. Consequently, we defined two distinct agent profiles for these roles. 

Agent PO understands the vision of the project. It is responsible for managing product backlog and prioritizing user stories based on business value and customer needs. This agent ensures that the user stories align with the overall product strategy and objectives.

Agent RE is tailored to focus on the quality aspects of user stories. It ensures that the user story description is unambiguous, and the acceptance criteria are measurable. This is crucial for verifying that the story fulfills its objectives upon implementation.

The construction of these agent profiles is an iterative process, encompassing role definition and expectation, key responsibilities, practical tips, and tone adjustment.

The role definition and expectation set an expectation for the agent’s performance in this role. It allows us to draw on the foundation of folk psychology concepts we use to understand human behavior, i.e. beliefs, desires, goals, ambitions, emotions, etc. \cite{shanahan2023role}. In the role play description, we set a high benchmark for the agent’s expected performance, akin to a human requirements engineer's knowledge but at an advanced level (level 250 compared to level 10). This highlights our expectations from the agent. The following example, quoted from the RE agent profile configuration, illustrates this. ``\textit{From now on, you will play the role of a Requirements Engineer, a new version of AI model that is capable of analyzing, documenting, and managing software requirements. If a human Requirements Engineer has a level 10 knowledge, you will have a level 250 of knowledge in this role. Please make sure to make accurate and comprehensive results in this role, because if you don't, the software may not meet the desired outcomes, and the project could fail. }''

The key responsibilities of each agent specify the critical requirements engineering tasks in software projects. Such descriptions provide guidance to ensure that agents follow a structured approach throughout the software development process, as shown in an example

``\textit{Your main task is to elicit, analyze, document, and manage the requirements for a software project.}''.

Practical tips are also provided to enhance the effectiveness of the agents in executing a specific task. An example for the RE agent is ``\textit{Use clear and unambiguous language when documenting requirements to avoid any misunderstandings}” This guidance is crucial for maintaining clarity and precision in requirements documentation.

Furthermore, we crafted the tone specification \cite{wu2022ai}\cite{zamfirescu2023johnny} to set additional expectations. For the PO agent, the tone is described as 
``\textit{The tone of the responses should be professional, yet approachable and friendly. As a Product Owner, you should provide clear and concise instructions, while also fostering a positive and collaborative environment.}
This ensures that the communication style of the agent is professional, precise, and objective, prioritizing clarity and conciseness.

In summary, the agent profiles are designed to reflect the real-life roles of a PO and RE in agile teams at Austrian Post Group IT. The agents are expected to not only understand their specific tasks but also execute them with a high level of expertise and in a manner that is conducive to collaborative software development.

\vspace{-10pt}
\begin{figure}[!ht]
\centering
\includegraphics[scale=0.8]{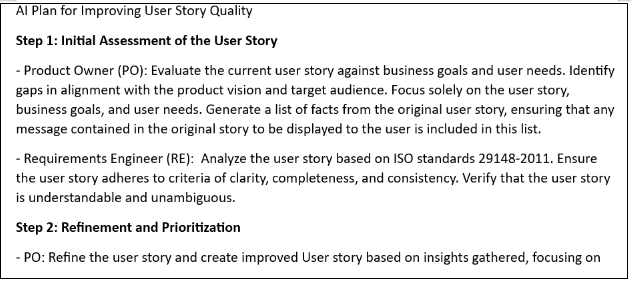}
\caption{An example excerpted from the generated AI plan}
\label{fig:AI plan example}
\end{figure}
\vspace{-20pt}

\subsubsection{Subtasks}
\begin{figure}[ht]
\centering
\includegraphics[width=\textwidth]{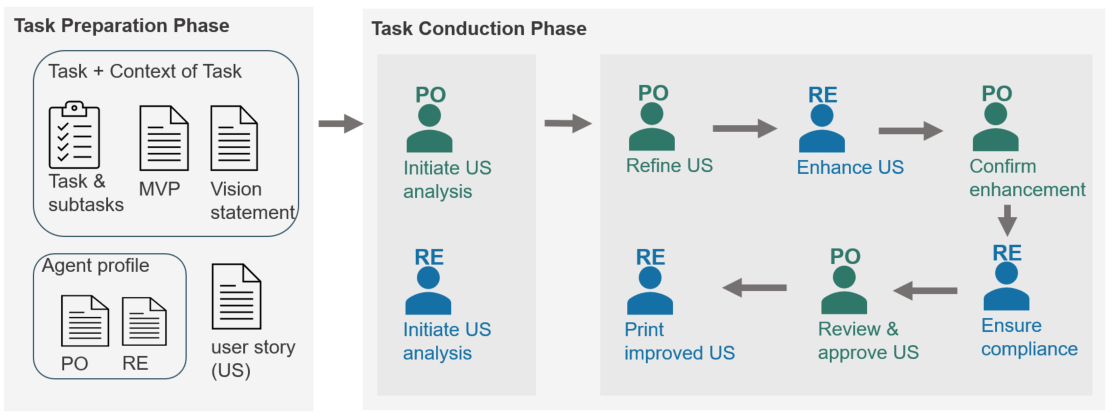}
\caption{AI plan illustrated in the task conduction phase}
\label{fig:AI plan}
\end{figure}

After specifying the task and identifying the participating agents, our next step involves detailing the sequence of interactions between these agents. To achieve this, we used an AI plan pattern \cite{silver2022pddl} to generate a comprehensive list of key steps and subtasks for task completion, as well as the identification of the responsible agents. Fig. \ref{fig:AI plan example} depicts part of the AI plan, illustrating the specific subtasks that Agents PO and RE are expected to perform as a first step in the task of enhancing user story quality. This plan was further reviewed and refined by a Scrum master and a PO in agile teams, ensuring that it aligns with the company's agile framework, common practice for requirements management, and project objectives. Fig. \ref{fig:AI plan} visualizes the complete structured conversation flow between the two agents and their subtasks in the task conduction phase, i.e. the collaborative and iterative interaction between agents in the user story improvement process.

\vspace{-15pt}
\subsubsection{GPT models}
A challenge encountered during our experiment was exceeding the token limit, particularly when agents engaged in response exchanges across various subtasks. After several adjustments of prompt composition, the gpt-3.5-turbo-16k model and the advanced gpt-4-1106-preview model were chosen. Both models are advanced iterations in OpenAI's GPT series \cite{OpenAImodels}. GPT-3.5-Turbo-16K is tailored for quicker response and can efficiently manage up to 16k tokens, making it suitable for extended conversations or documents. The GPT-4-1106-Preview further advances the language model capabilities to handle 128k tokens and returns a maximum of 4096 output tokens. 

\subsection{Evaluation}
When the experiment was set up, ALAS was ready for the task of improving the quality of user stories for the mobile delivery application. To evaluate how effectively ALAS accomplishes the task, we surveyed professionals from six agile teams at the Austrian Post Group IT. The survey focuses on gathering feedback from professionals to assess the effectiveness, benefits, potential concerns, and overall satisfaction with ALAS's improved user stories. We prepared a questionnaire based on the characteristics of good requirements specified in the INVEST framework \cite{INVEST}. The statements are shown in Table \ref{questionnaire statements}.

Participants evaluated user stories using a Likert scale from 1 to 5, where 1 indicates strong disagreement and 5 indicates strong agreement. Additionally, the survey includes two open-ended questions to collect participants’ feedback on specific improvements made to the original user stories, concerns about the improved versions, and suggestions for further improvements. Finally, participants will provide an overall satisfaction rating for the improved user stories, and identify those most appropriate for the project context.

\begin{table}[!ht]
\centering
\caption{Characteristics of good user stories used in the questionnaire}
\label{questionnaire statements}
\begin{tabular}{p{1cm} p{11cm}}
\hline
\textbf{ID} & \textbf{Statement} \\ 
\hline
S1 & The user story is simple and easy to understand.\\	
S2 & The user story is of the right size (not too long). \\	
S3 & The user story is at a suitable level of detail. \\
S4 & The user story includes a description of the task and the goal to achieve. \\
S5 & The user story is technically achievable. \\
S6 & The acceptance criteria include measurable elements for test case preparation. \\
S7 & The acceptance criteria are sufficient to validate the story.\\
\hline
\end{tabular}
\end{table}

Considering the time required for participants to complete the survey, the questionnaire includes just two original user stories and two improved versions of each. 
Fig. \ref{fig:userstoryexample} depicts one of the two original user stories, i.e. user story 1 (US1). For each improved version, participants' assessment is based on seven statements in Table \ref{questionnaire statements} and the answer to corresponding open-ended questions. The survey\footnote{http://tinyurl.com/4veet5me} included %two user stories and their improved versions, 
a total of 34 rating questions, and 12 open-ended questions. By analyzing both quantitative and qualitative feedback from the participants, we aim to validate the effectiveness of ALAS and identify opportunities for further enhancements in our approach.

\section{Results}

Our questionnaire gathered 12 responses from six agile teams. The participants include two POs, four developers, a test manager, a Scrum master, a requirements analyst, two testers, and a train coach. Notably, 10 out of 11 participants have been working at the company for over two years, and 9 have more than five years of extensive experience in agile projects. Their expertise provided a solid foundation for evaluating the user stories presented in the survey. Upon a brief introduction to the survey's objectives, participants were fully engaged and dedicated an average of 33 minutes to complete the questionnaire.

The survey participants reported their concerns about User Stories 1 and 2 (US1 and US2). Both stories were mainly criticized for their ambiguity and lack of essential details, especially in the AC which failed to describe the necessity of certain conditions that were described as criteria. In addition, the business value in these descriptions remains vague. Specific scenarios, especially those involving error handling in US1 are also inadequately addressed.

The improved versions, US1(v.1) and US2(v.1), produced by GPT-3.5-turbo agents, exhibited significant improvements in clarity and comprehensibility. They enhanced the clarity of AC and improved the narrative flow, leading to a more coherent presentation of user stories. However, feedback from survey participants highlighted that the new titles for user stories were too creative and the description in AC should be more detailed. In addition, concerns remain in the AC about scenarios such as multiple printer connections identified in US1.  

Improvements produced by the GPT-4 model, i.e. US1(v.2) and US2(v.2), were recognized for their comprehensive content and clearer expression of the business value of the stories. Specifically, the AC in US1(v.2) is formulated more clearly and completely, addressing printer connection problems which were ambiguous in US1 and US1(v.1). Nonetheless, this increase in detail and clarity led to a significant increase in story complexity and length, which could potentially undermine their practical applicability – six survey participants noted concerns about user story descriptions being too long.

%\vspace{-10pt}
\begin{figure}[ht]
\centering
\includegraphics[width=\textwidth]{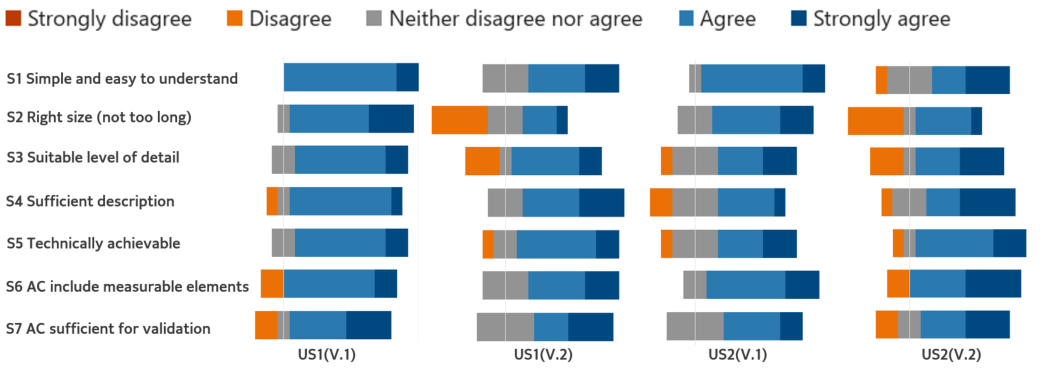}
\caption{Distribution of survey participants' perceptions of user story quality, Note: US* = User Story *, v.1 = Version 1 Improved by gpt-3.5-Turbo, v.2 = Version 2 Improved by gpt-4-1106-Preview}
\label{fig: rating distribution}
\end{figure}

%\vspace{-30pt}

\begin{table}[!ht]
    \centering
    \caption{Average ratings (1-5 Scale) of overall satisfaction and quality of user stories. }
    \label{tab:user_stories}
    \begin{tabular}{p{2cm} p{1.2cm} p{1.2cm} p{1.2cm} p{1.2cm} p{1.2cm} p{1.2cm} p{1.2cm} p{1.2cm}}
    \hline
    \textbf{User story} & \textbf{S1} & \textbf{S2} & \textbf{S3} & \textbf{S4} & \textbf{S5} & \textbf{S6} & \textbf{S7} & \textbf{Overall rating} \\ 
    \hline
    US1 & - & - & - & - & - & - & - & 3.33 \\ 
    US1(v.1) & 4.17 & 4.25 & 4 & 3.83 & 4 & 3.83 & 3.92 & \textbf{4} \\
    US1(v.2) & 3.92 & \underline{3} & 3.58 & 4.08 & 3.83 & 3.92 & 3.92 & \textbf{4} \\
    US2 & - & - & - & - & - & - & - & 2.79 \\
    US2(v.1) & 4.08 & 4 & 3.75 & 3.42 & 3.75 & 4.08 & 3.75 & 3.54 \\
    US2(v.2) & 3.83 & \underline{3.17} & 3.75 & 4 & 4 & 4.08 & 3.8 & \textbf{3.71} \\
    \hline
    \end{tabular}
\end{table}

Additionally, the user story quality satisfaction ratings complement the insights from participants' feedback, as illustrated in Fig. \ref{fig: rating distribution} which presents a distribution of participants' perceptions of user story quality and Table \ref{tab:user_stories} which summarizes average ratings for overall satisfaction and quality aspects of the user stories. Both US1(v.1) and US1(v.2) scored an average overall satisfaction of 4, while US2(v.2) scored 3.71, higher than US2(v.1). This preference is confirmed by 7 participants choosing US1(v.2) and US2(v.2) for the project. However, despite their merit on sufficient description (S4), both USs rated lower in simplicity, brevity, and appropriate level of detail (S1, S2, and S3), particularly  struggling with their size, scoring averages of 3 and 3.17 respectively. Notebly, US2(v.2) received the most disagreements regarding its size, with 5 participants marking "Disagree". This disparity may potentially affect the user story’s comprehensibility(S1). US1(v.2) also received a minor drop in technical achievability (S5), compared to US1(v.1). These results highlight concerns over the increased length and complexity of user stories generated by the GPT-4 model, significantly affecting the satisfaction level of these user stories, a sentiment corroborated by the survey results.

\section{Discussion}

Our experiments with ALAS for user story quality improvement have demonstrated significant benefits in enhancing user story quality, particularly in terms of clarity, specificity, and business value articulation. This is evident from the increased overall satisfaction ratings given by survey participants. These findings indicate that ALAS effectively refines user stories for improved quality.

Despite these enhancements, agents’ ability to learn from context, while impressive, highlights a gap in aligning with project-specific contexts and requirements. One developer's feedback in the survey noted that US1(v.2) included an authentication process that, while relevant to the story, \textit{"seems to be out of scope of the US1"}. Similar feedback was observed from another developer’s feedback. These imply that some quality aspects of requirements may be missing or unclearly specified in the agent prompts related to their responsibilities. Consequently, careful prompt preparation and rigorous evaluation by human experts, such as the PO, are essential. When implementing ALAS for specific tasks, engaging the PO and domain experts during the task preparation phase becomes crucial to optimize prompts for the desired output.

Considering that we have only two agents, PO and RE, integrated into ALAS, we can explore incorporating additional specialized agents, such as a tester agent to check factual information and refine acceptance criteria. Similarly, a quality analyst agent could monitor the scope, level of detail, and relevance of the story description, ensuring focus and preventing scope creep, mirroring agile project practices. Currently, ALAS's outputs require manual validation by the Product Owner (PO) to align with project goals and stakeholder expectations. This manual validation is crucial to mitigate the limitations of automated generation and preserve the practical utility of user stories.

In examining the parameters governing GPT models, particularly the 'Temperature' parameter that stimulates creativity, we observe a double-edged sword. While it boosts novel and diverse content generation, it also increases the risk of AI hallucination \cite{rawte2023troubling}, which can lead to plausible yet inaccurate or irrelevant outputs. Addressing AI hallucination necessitates careful parameter tuning to ensure that harnessed creativity enhances rather than detracts from user story quality. In our experiments, we set the medium value 1 for Temperature. However, this still poses a challenge in maintaining factual accuracy, emphasizing the need for an integrated role that guides and monitors the overall discussion.

In summary, while ALAS has made significant strides in improving user story quality, there remains room for further refinement. The integration of specialized agents, parameter optimization, and human expertise will contribute to an evolving ALAS implementation. Future improvements should focus on enhancing contextual alignment, refining acceptance criteria specificity, and mitigating the risk of irrelevant or inaccurate content generation. By addressing these areas, ALAS can become a more robust tool, bridging the gap between automated language generation and the nuanced demands of software development narratives.

\section{Conclusion}

In this study, we presented a reference model for an agent-based system that utilizes LLMs as agents to aid software development tasks. The reference model guided the implementation of ALAS, which integrates GPT models as agents to enhance requirement quality in agile software development. The experimental results demonstrated that ALAS significantly improves user story clarity, comprehensibility, and alignment with business objectives. However, the findings also underscored the indispensable role of human intelligence, particularly the PO in software projects, who facilitate and monitor the improvements in user stories to guarantee the integrity of automatically produced outputs. Moving forward, enhancing ALAS necessitates not only incorporating specialized agents with optimized profiles and task descriptions but also fine-tuning AI parameters to minimize hallucinations and enhance contextual accuracy. This paper contributes a foundational framework and a proof-of-concept for AI-assisted user story quality improvement, marking a significant step forward in bridging the gap between AI capabilities and human expertise in software development.

\bibliographystyle{splncs04}
\bibliography{LLM_agent}

\begin{thebibliography}{10}
\providecommand{\url}[1]{\texttt{#1}}
\providecommand{\urlprefix}{URL }
\providecommand{\doi}[1]{https://doi.org/#1}

\bibitem{6146379}
Iso/iec/ieee international standard - systems and software engineering -- life cycle processes --requirements engineering. ISO/IEC/IEEE 29148:2011(E) pp. 1--94 (2011). \doi{10.1109/IEEESTD.2011.6146379}

\bibitem{INVEST}
Invest in good stories, and smart tasks. \url{https://xp123.com/articles/invest-in-good-stories-and-smart-tasks/} (Accessed: 2024-1-10)

\bibitem{OpenAImodels}
Openai. \url{https://platform.openai.com/docs/models} (Accessed: 2024-1-10)

\bibitem{NABC}
Sri international best practice. \url{https://web.stanford.edu/class/educ303x/wiki-old/uploads/Main/SRI_NABC.doc} (Accessed: 2024-1-10)

\bibitem{berry2006new}
Berry, D.M., Bucchiarone, A., Gnesi, S., Lami, G., Trentanni, G.: A new quality model for natural language requirements specifications. In: Proceedings of the international workshop on requirements engineering: foundation of software quality (REFSQ) (2006)

\bibitem{brown2020language}
Brown, T., Mann, B., Ryder, N., Subbiah, M., Kaplan, J.D., Dhariwal, P., Neelakantan, A., Shyam, P., Sastry, G., Askell, A., et~al.: Language models are few-shot learners. Advances in neural information processing systems  \textbf{33},  1877--1901 (2020)

\bibitem{cohn2004user}
Cohn, M.: User stories applied: For agile software development. Addison-Wesley Professional (2004)

\bibitem{dalpiaz2019detecting}
Dalpiaz, F., Van Der~Schalk, I., Brinkkemper, S., Aydemir, F.B., Lucassen, G.: Detecting terminological ambiguity in user stories: Tool and experimentation. Information and Software Technology  \textbf{110},  3--16 (2019)

\bibitem{fabbrini2001linguistic}
Fabbrini, F., Fusani, M., Gnesi, S., Lami, G.: The linguistic approach to the natural language requirements quality: benefit of the use of an automatic tool. In: proceedings 26th annual NASA Goddard software engineering workshop. pp. 97--105. IEEE (2001)

\bibitem{ferreira2022towards}
Ferreira, A.M., da~Silva, A.R., Paiva, A.C.: Towards the art of writing agile requirements with user stories, acceptance criteria, and related constructs. In: ENASE. pp. 477--484 (2022)

\bibitem{glinz2020handbook}
Glinz, M., van Loenhoud, H., Staal, S., B{\"u}hne, S.: Handbook for the cpre foundation level according to the ireb standard. International Requirements Engineering Board  (2020)

\bibitem{lucassen2015forging}
Lucassen, G., Dalpiaz, F., Van Der~Werf, J.M.E., Brinkkemper, S.: Forging high-quality user stories: towards a discipline for agile requirements. In: 2015 IEEE 23rd international requirements engineering conference (RE). pp. 126--135. IEEE (2015)

\bibitem{lucassen2016improving}
Lucassen, G., Dalpiaz, F., van~der Werf, J.M.E., Brinkkemper, S.: Improving agile requirements: the quality user story framework and tool. Requirements engineering  \textbf{21},  383--403 (2016)

\bibitem{nguyen2023generative}
Nguyen-Duc, A., Cabrero-Daniel, B., Przybylek, A., Arora, C., Khanna, D., Herda, T., Rafiq, U., Melegati, J., Guerra, E., Kemell, K.K., Saari, M., Zhang, Z., Le, H., Quan, T., Abrahamsson, P.: Generative artificial intelligence for software engineering--a research agenda. arXiv preprint arXiv:2310.18648  (2023)

\bibitem{austriaPost}
Niessl, M., Gruber, C., Eder, M.: Restarting scaled agile development at austrian post. Experience Report, 24th International Conference on Agile Software Development  (2023)

\bibitem{rawte2023troubling}
Rawte, V., Chakraborty, S., Pathak, A., Sarkar, A., Tonmoy, S., Chadha, A., Sheth, A.P., Das, A.: The troubling emergence of hallucination in large language models--an extensive definition, quantification, and prescriptive remediations. arXiv preprint arXiv:2310.04988  (2023)

\bibitem{ronanki2023investigating}
Ronanki, K., Berger, C., Horkoff, J.: Investigating chatgpt’s potential to assist in requirements elicitation processes. In: 2023 49th Euromicro Conference on Software Engineering and Advanced Applications (SEAA). pp. 354--361. IEEE (2023)

\bibitem{shanahan2023role}
Shanahan, M., McDonell, K., Reynolds, L.: Role play with large language models. Nature  \textbf{623}(7987),  493--498 (2023)

\bibitem{silver2022pddl}
Silver, T., Hariprasad, V., Shuttleworth, R.S., Kumar, N., Lozano-P{\'e}rez, T., Kaelbling, L.P.: Pddl planning with pretrained large language models. In: NeurIPS 2022 foundation models for decision making workshop (2022)

\bibitem{wei2022chain}
Wei, J., Wang, X., Schuurmans, D., Bosma, M., Xia, F., Chi, E., Le, Q.V., Zhou, D., et~al.: Chain-of-thought prompting elicits reasoning in large language models. Advances in Neural Information Processing Systems  \textbf{35},  24824--24837 (2022)

\bibitem{white2023chatgpt}
White, J., Hays, S., Fu, Q., Spencer-Smith, J., Schmidt, D.C.: Chatgpt prompt patterns for improving code quality, refactoring, requirements elicitation, and software design. arXiv preprint arXiv:2303.07839  (2023)

\bibitem{wiegers2013software}
Wiegers, K., Beatty, J.: Software requirements. Pearson Education (2013)

\bibitem{wu2022ai}
Wu, T., Terry, M., Cai, C.J.: Ai chains: Transparent and controllable human-ai interaction by chaining large language model prompts. In: Proceedings of the 2022 CHI conference on human factors in computing systems. pp. 1--22 (2022)

\bibitem{zamfirescu2023johnny}
Zamfirescu-Pereira, J., Wong, R.Y., Hartmann, B., Yang, Q.: Why johnny can’t prompt: how non-ai experts try (and fail) to design llm prompts. In: Proceedings of the 2023 CHI Conference on Human Factors in Computing Systems. pp. 1--21 (2023)

\bibitem{zhang2022automatic}
Zhang, Z., Zhang, A., Li, M., Smola, A.: Automatic chain of thought prompting in large language models. arXiv preprint arXiv:2210.03493  (2022)

\end{thebibliography}

\end{document}